%% file: paper.tex
\documentclass[10pt,aps,prb,twocolumn,showpacs,a4paper]{revtex4-1}

\usepackage{amsmath}   
\usepackage{mathtools} 
\usepackage{amssymb}   
\usepackage{mathrsfs}  
\usepackage{textcomp}  
\usepackage{accents}   

\usepackage{graphicx}  
\usepackage{epstopdf}  
\usepackage{subfigure} 

\usepackage{makecell}
\usepackage{tabu}
\usepackage{array}     
\usepackage{tabularx}  
\usepackage{multirow}  
\usepackage{booktabs}  
\usepackage{dcolumn}   

\usepackage{gensymb}   

\usepackage{setspace}

\usepackage[usenames,dvipsnames]{color} 
\usepackage{url}       
\usepackage{hyperref}  

\hypersetup{
  colorlinks,
  citecolor=Violet,
  linkcolor=Red,
  urlcolor=Blue}

\newcommand{\kelvin}{$\,$K}
\newcommand{\degreeC}{$\,$\celsius{}}

\definecolor{myred}{rgb}{1,0,0}
\definecolor{mygreen}{rgb}{0,0.7,0}
\definecolor{myblue}{rgb}{0,0,1}
\definecolor{mypink}{rgb}{0.8,0,0.8}
\definecolor{darkred}{rgb}{0.6,0,0}

\newcommand{\red}[1]    {\color{myred}#1}

\graphicspath{{./figures/}}

\begin{document}
\setlength{\heavyrulewidth}{0.08em}
\setlength{\lightrulewidth}{0.05em}
\setlength{\cmidrulewidth}{0.03em}
\setlength{\belowrulesep}{0.65ex}
\setlength{\belowbottomsep}{0.00pt}
\setlength{\aboverulesep}{0.40ex}
\setlength{\abovetopsep}{0.00pt}
\setlength{\cmidrulesep}{\doublerulesep}
\setlength{\cmidrulekern}{0.50em}
\setlength{\defaultaddspace}{0.50em}
\setlength{\tabcolsep}{4pt}

\title{The Role of Temperature and Magnetic Effects on the Stacking-fault Energy in Austenitic Iron}

\author{Seyyed Arsalan Hashemi}
\email{arsalan.hashemi@ph.iut.ac.ir}
\affiliation{Department of Physics, Isfahan University of Technology, Isfahan, Iran}

\author{Hojjat Gholizadeh}
\email{h.gholizadeh@ph.iut.ac.ir}
\affiliation{Department of Physics, Isfahan University of Technology, Isfahan, Iran}

\author{Hadi Akbarzadeh}
\email{akbarzad@cc.iut.ac.ir}
\affiliation{Department of Physics, Isfahan University of Technology, Isfahan, Iran}

\date{\today}

\begin{abstract}
We have investigated the role of temperature and magnetic effects on the stacking-fault energy (SFE) in pure austenitic iron based on density functional theory (DFT) calculations. 
Using the axial next-nearest-neighbor Ising (ANNNI) model, the SFE is expanded in terms of free energies of bulk with face-centered cubic (fcc), hexagonal close-packed (hcp), and double-hcp (dhcp) structures. 
The free-energy calculations require the lattice constant and the local magnetic moments at various temperatures. 
The earlier is obtained from the available experimental data, while the later is calculated by accounting for the thermal magnetic excitations using Monte-Carlo techniques. 
Our results demonstrate a strong dependence of the SFE on the magnetic effects in pure iron. 
Moreover, we found that the SFE increases with temperature. 
%
%
\end{abstract}

\maketitle


\input{introduction}

\input{methodology}

\input{conclusion}

\begin{acknowledgments}
...
\end{acknowledgments}

\bibliographystyle{apsrev4-1}
\bibliography{hgh}
\end{document}

%% file: introduction.tex
\section{Introduction}
\label{sec:introduction}

Iron and its alloys have played a significant role in the development of human civilization. 
Beside its industrial importance, some unique features of iron like its phase transitions and magnetic properties have opened interesting fields of research for materials scientist.

The well-known phase diagram of iron shows that, at atmospheric pressure, and at low temperatures, pure iron is found in ferromagnetic body-centered cubic (bcc) structure \cite{Massalski1986}. 
As temperature rises, at 1043\kelvin{} (770\degreeC{}), \textit{i.e.}, at the Curie temperature of iron, it demonstrates a magnetic phase transition from the ferromagnetic to paramagnetic, while preserving the bcc structure. 
At 1185\kelvin{} (912\degreeC{}), iron faces a structural phase transition from the bcc to the face-centered cubic (fcc) structure. 
Further heating of iron reveals a second structural phase transition at 1667\kelvin{} (1394\degreeC{}), through which the fcc structure changes back to the bcc. 
Finally, at 1811\kelvin{} (1538\degreeC{}) iron melts \cite{Massalski1986}. 
Although the pure iron is found in the bcc structure at room temperatures, the addition of alloying elements like manganese and nickel can stabilize its fcc phase at room temperature \cite{Massalski1986,McGuire2008}.

The mechanical properties of steels is influenced by their plastic deformations.
In fcc metals, plastic deformations may occur through different mechanisms including dislocation gliding, twinning (twinning-induced plasticity, TWIP), and phase transformation (transformation-induced plasticity, TRIP). 
The activation of these mechanisms has been proven to be governed by the stacking-fault energy (SFE) \cite{Frommeyer2003,Grassel2000,Allain2004a,Dumay2008}.

\begin{figure}[thb]
  \centering
  \includegraphics[width=0.850\columnwidth]{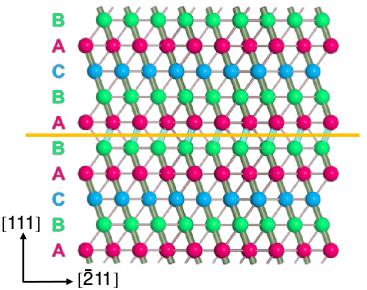}
  \caption{The stacking sequence of an fcc structure along its [111] direction, with an intrinsic stacking fault at the position denoted by the horizontal orange line. 
           The perfect fcc stacking order is highlighted in green, while the interruption due to the fault is emphasized by the turquoise color.}
  \label{fig:ISF_Side_View}%
\end{figure}

The intrinsic stacking fault (SF) is one of the simplest planar defects in the fcc crystal lattice \cite{Hirth1982}. 
The fcc structure is formed by the stacking of the close-packed layers in the $\ldots ABCABCABC\ldots$ sequence. 
In this structure, an ISF can be considered as the elimination of a close-packed layer in the bulk, resulting in the $\ldots ABC\mathbf{AB}|\mathbf{AB}C\ldots$ sequence, where the removed layer has been denoted by a vertical line. 
This defect is demonstrated in Fig.~\ref{fig:ISF_Side_View}. 
In the neighborhood of the fault, the structure resembles locally the stacking sequence of the hexagonal close-packed (hcp) structure, as highlighted by bold letters in the notation \cite{Hirth1982,Hammer1992} (see Fig.~\ref{fig:ISF_Side_View}). 
The energy associated with a SF, the SFE, is defined as the difference between the free energy of a structure with a fault and that of the perfect fcc structure: ${\mathrm{SFE}}=F_{\mathrm{SF}}-F_{\mathrm{fcc}}$.

There has been numerous experimental works on the measuring of the SFE in different austenitic steels \cite{Reed1974,Adeev1975,Bampton1978}. 
As already discussed by Abbasi \cite{Abbasi2011}, Gholizadeh \cite{Gholizadeh2013}, and Reyes-Huamantinco \cite{Reyes-Huamantinco2012}, the available experimental data for the SFE are highly questionable since the reported ranges are too broad and have shown a high deviation from average amount. 
Therefore, the development of a theoretical approach based on quantum mechanics is highly desired.

In recent years, the SFE and its dependence on different parameters have been the subject of many theoretical investigations within the framework of the density functional theory (DFT) \cite{Hohenberg1964,Kohn1965a}. 
Some of them have applied supercell approaches, which allow for the explicit simulation of the fault as well as for the relaxation of local forces, but critically restricts the calculation of magnetic or chemical disorders. 
For instance, Kibey \textit{et al.} \cite{Kibey2006} have studied the dependence of the SFE on the concentration of interstitial nitrogen in fcc iron. 
Similarly, Abbasi \textit{et al.} \cite{Abbasi2011} and Gholizadeh \textit{et al.} \cite{Gholizadeh2013} have studied the influence of interstitial carbon of the SFE in fcc iron. 
In these three works, the lattice local displacements introduced by the interstitial atom and also by the fault are accurately accounted for. 
On the other hand, using spin-unpolarized simulations according to 0\kelvin{} equilibrium, all magnetic interactions in the paramagnetic medium are neglected. 
Abbasi \cite{Abbasi2011} reports only small differences between results obtained from tests with nonmagnetic (spin-unpolarized) and ferromagnetic calculations, supporting their simplified nonmagnetic calculations where the influence of the magnetic interactions on the qualitative behavior of the SFE is assumed negligible. 
Referring to Abbasi's tests, Gholizadeh \cite{Gholizadeh2013} states that although a nonmagnetic calculation may be reliable enough to study the dependence of the SFE against the interstitial concentration, calculating the absolute value of the SFE and also developing a complete understanding of the atomic interactions in the paramagnetic medium requires accurately accounting for the magnetic interactions.

Other investigations have applied Green's function formalism \cite{Korringa1947,Kohn1954,Zabloudil2004,Ruban2008}, where the utilization of the disordered local moments (DLM) \cite{Gyorffy1985} approach and the coherent potential approximation (CPA) \cite{Soven1967,Taylor1967,Gyorffy1972} allows for the simulation of the magnetic and chemical disorders, respectively. 
For instance, Vitos \textit{et al.} \cite{Vitos2006} studied the temperature dependence of the SFE in iron--chromium--nickel alloys. 
Vitos concludes that the temperature dependence of the SFE is almost totally dictated by the contribution of the magnetic fluctuations into the free energy. 
Later, Reyes-Huamantinco \textit{et al.} \cite{Reyes-Huamantinco2012} and Gholizadeh \textit{et al.} \cite{Gholizadeh2013a} improved Vitos's approach, particularly by including the experimental data for the thermal lattice expansion, and calculated the temperature dependence of the SFE in iron-manganese and iron--chromium--nickel alloys, respectively. 
The two works reveal that the temperature dependence of the SFE is mainly obtained from the total energy of the alloy, which is in turn a function of its lattice parameter at different temperatures. 
Therefore, in contrast with Vitos's results, the two works conclude that the temperature dependence of the SFE is mainly dictated by the lattice thermal expansion. 

Although the published works emphasize that accounting for magnetic effects is crucial for understanding the phase stability and hence the behavior of the SFE in Fe-based alloys, an explicit comparison between quantitative results obtained from the paramagnetic calculations and those obtained from nonmagnetic (spin-unpolarized) calculations can reveal the magnitude of the magnetic interactions in the SFE. 
Iron is the dominant element in many industrially interested alloys, including those mentioned above. 
Further investigations show that the Fe atoms are the main responsible for the magnetic interactions in these alloys \cite{Reyes-Huamantinco2012,Gholizadeh2013a}. 
Moreover, performing calculations for pure iron avoids all complexities which are related to an alloy compared to an element, like atomic size mismatch, local lattice distortion, short-range and long-range orders, Suzuki effect, chemical phase transitions \textit{etc}. 
Therefore, in the current work we compare two sets of the SFEs calculated for pure iron in fcc phase, one set using the methodology introduced by Reyes-Huamantinco \textit{et al.} \cite{Reyes-Huamantinco2012}, and the other set using nonmagnetic calculations. 
Such comparison will answer two main questions: 
(i) How does the SFE change with temperature in the paramagnetic austenitic iron? 
and, (ii) How big is the influence of the magnetic effects on the SFE in the paramagnetic austenitic iron?

%% file: methodology.tex
\section{Methodology}
\label{sec:methodology}

\subsection{The ANNNI Model}
\label{subsec:methodology:ANNNI}
\begin{figure}[thb]
  \centering
  \includegraphics[width=1.0\columnwidth]{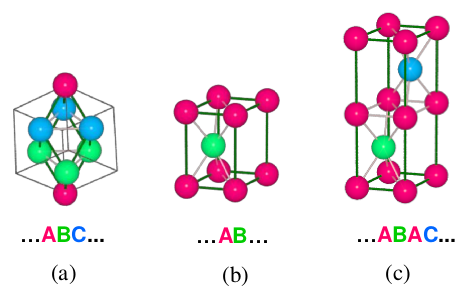}
  \caption{Primitive cells of three crystal structures used in the ANNNI model. 
     Atoms are colored according to their stacking position along the [111] direction.  
     {\red{(a)}} depicts the primitive cell of the fcc structure with only one atomic site. 
     The cubic cell is shown for a better imagination of the lattice. 
     {\red{(b)}} shows the primitive cell of the hcp structure containing two non-equivalent atoms (two atomic sites). 
     {\red{(c)}} represents the primitive cell of the dhcp structure with four non-equivalent atoms. }
  \label{fig:Unit_Cells}%
\end{figure}
The axial next-nearest-neighbor Ising (ANNNI) model, as explained by Cheng \textit{et al.} \cite{Cheng1987} and Denteneer \textit{et al.} \cite{Denteneer1987}, expands the SFE in terms of the free energies of bulk unit-cells with different structures. 
In its second order, the ANNNI model results in
\begin{equation}
   \mathrm{SFE}(T)=\frac{F^{\mathrm{hcp}}(T)+2F^{\mathrm{dhcp}}(T)-3F^{\mathrm{fcc}}(T)}{A}+\mathcal{O}(3),
   \label{eq:ANNNI_SFE}
\end{equation}
where $F^{\phi}$ denotes the Helmholtz free energy of a single atom in phase $\phi$, $A$ is the area in a close-packed layer occupied by a single atomic site, \textit{i.e.}, $A=\frac{\sqrt{3}}{4}a_{\mathrm{fcc}}^2$, and $\mathcal{O}(3)$ is the error introduced by neglecting the higher order interactions. 

\subsection{Temperature Dependence of the Free Energy}
\label{subsec:methodology:free_energy}

The Helmholtz free energy is defined as 
\begin{equation}
   F(T)=E(T)-TS(T),
   \label{eq:Helmholtz_free_energy}
\end{equation}
where $E(T)$ and $S(T)$ are the total (internal) energy and entropy, respectively.

\begin{figure}[ht]
\centering
\includegraphics[scale=0.45]{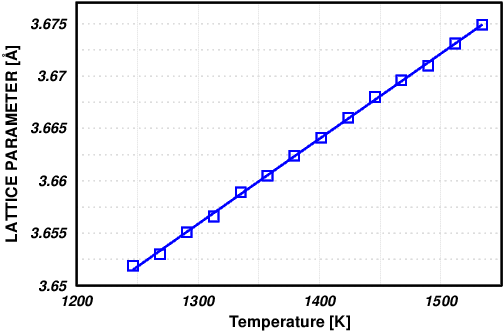}
\caption{The experimental lattice parameter of pure iron as a function of temperature, obtained from high-temperature XRD measurements \cite{Seki2005}.}
\label{seki}
\end{figure}

The total energy $E(T)$ is calculated using the DFT \cite{Hohenberg1964,Kohn1965a}. 
Its temperature dependence originates from three sources: (i) the electron distribution over states defined by the set of occupation numbers $\{\alpha_{\epsilon}\,|\,\epsilon\!\le\!\epsilon_{\mathrm{F}}\}$, (ii) the lattice parameter $a(T)$, and (iii) the average local magnetic moment $m(T)$:
\begin{equation}
   E(T)=E\big(\{\alpha_{\epsilon}\},a(T),m(T)\big).
   \label{eq:E(T)}
\end{equation}
The Mermin functional \cite{Mermin1965}, is applied in finite-temperature DFT calculations to account for the temperature dependence of the electron distribution of over states \cite{Martin2004}. 
The temperature dependence of the lattice parameter is taken into account by selecting the lattice parameters according to the experimental data for thermal lattice expansions (see Fig.~\ref{seki}). 
Finally, the temperature dependent local magnetic moments are evaluated using a statistical approach, which is explained in the next subsection.

\subsection{Longitudinal Spin Fluctuations}
\label{subsec:methodology:LSF}

The thermal excitation of the local magnetic moments in the paramagnetic DLM state is described using a simple model based on the theory of unified itinerant magnetism \cite{Moriya1985}. 
Originally, the theory accounts for both transverse and longitudinal spin fluctuations on equal footing. 
However, here it is further simplified so that the transverse spin fluctuations, \textit{i.e.}, the fluctuations in the orientations of the magnetic moments, always follow the completely disordered configuration described in the DLM state. 
The longitudinal spin fluctuations (LSF), \textit{i.e.}, the fluctuations in the size of the magnetic moments, are obtained by performing a classical Monte-Carlo simulation over a mapping of the system energetics. 
These system energetics are calculated using the EMTO code. 

For representing the energy of the classical magnetic state, Ruban \textit{et al.} \cite{Ruban2007} introduced a magnetic Hamiltonian which was later applied to an Fe--Mn alloy by Reyes-Huamantinco \textit{et al.} \cite{Reyes-Huamantinco2012}. 
In the case of pure iron, the Hamiltonian is simplified as
\begin{equation}
   H_{\mathrm{mag.}}(m)=J(m),
   \label{eq:LSF_Reyes}
\end{equation}
where $m$ is the \emph{spatially-averaged} local magnetic moments of iron atoms, and $J(m)$ is the energy required to excite this averaged moment from 0 to the value $m$ in the DLM paramagnetic state. 
After calculating the Hamiltonian parameters, $J(m)$, a Metropolis Monte-Carlo technique is applied to find the temperature-dependence of the local magnetic moments \cite{Ruban2007}.

\subsection{Entropy Contributions}
\label{subsec:methodology:entropy}

In a solid, the entropy consists of configurational, vibrational, magnetic, and electronic contributions \cite{Walle2002b,Ruban2008}: 
\begin{equation}
   S=S_{\mathrm{vib.}}+S_{\mathrm{conf.}}+S_{\mathrm{mag.}}+S_{\mathrm{el.}}, 
   \label{eq:entropy_contrinutions}
\end{equation}
where all terms are associated to a single site in the lattice.

Currently, there are no available theoretical tools to determine the vibrational entropy in paramagnetic random alloys. 

The configurational entropy is connected to the disorder in the arrangement of atoms of different species in the material. 
Therefore, in a pure element where all atoms are equal, it simply vanishes: $S_{\mathrm{conf.}}=0$.

For the ideal paramagnetic state, the magnetic entropy is evaluated using the mean-field expression \cite{Grimvall1989}
\begin{equation}
   S_{\mathrm{mag.}}(T)=k_{\mathrm{B}}\ln\big(m(T)+1\big).
   \label{eq:S_mag}
\end{equation}
Here, $m(T)$ denotes the average (over time and space) local magnetic moment obtained using the mentioned Monte Carlo calculations. 

The electronic entropy can be calculated as \cite{Vitos2007}
\begin{eqnarray}
   S_{\mathrm{el.}}(T)&=&-2k_{\mathrm{B}}\int\!\Big\{f(\epsilon)\ln\big(f(\epsilon)\big)\\
                      &+&\big(1-f(\epsilon)\big)\ln\big(1-f(\epsilon)\big)\Big\}D(\epsilon)\,\mathrm{d}\epsilon\nonumber,
\end{eqnarray}
where, $D(\epsilon)$ and $f(\epsilon)$ denote the density of states and the finite-temperature Fermi function \cite{Fermi1926}, respectively. 
The Fermi function is a consequence of the Fermi-Dirac statistics \cite{Fermi1926,Dirac1926} and gives the probability of occupation of a state with energy $\epsilon$ at temperature $T$. 
The Fermi function is defined as
\begin{equation}
   f(\epsilon)=\frac{1}{e^{(\epsilon-\mu)/k_{\mathrm{B}}T}+1},
\end{equation}
where $\mu$ indicates the chemical potential. 

\subsection{The SFE Calculations}
\label{subsec:methodology:SFE}

In order to find the total energy as a function of temperature, the temperature-dependent lattice parameter and local magnetic moments are used in a set of \emph{constrained} DFT calculations, where the magnetic moment is fixed to its corresponding value evaluated using the Monte-Carlo method. 
The free energy is simply found by subtracting the entropy contributions from the total energy (see Eqs.~\ref{eq:Helmholtz_free_energy} and \ref{eq:entropy_contrinutions}). 
The SFE is evaluated using these total energies, according to the Eq.~\ref{eq:ANNNI_SFE}.

%% file: conclusion.tex
\section{conclusion}
\label{sec:conclusion}

Our results evaluated using both spin-unpolarized and spin-polarized calculations show that the SFE increases with temperature. 
Although, spin-polarized calculations suggest that the temperature dependence of the SFE vanishes at high temperatures. 
The similar trend of the SFE in both calculations shows that the thermal dependence of the SFE is mainly originated from the thermal lattice expansion. 
This is in contrast with the published work by Vitos \textit{et al.} \cite{Vitos2006}, but agrees with results found by Reyes-Huamantinco \cite{Reyes-Huamantinco2012} and Gholizadeh \cite{Gholizadeh2013a}. 
We also find that, in order to get the correct phase stability of iron and also the correct value of its SFE, one must account for magnetic interactions in the system, including the spin fluctuations.